\documentclass[5p,twocolumn,times,number]{elsarticle}

\usepackage{graphicx}
\usepackage{amsmath} 

\usepackage{lineno}

\begin{document}

\begin{frontmatter}

\title{Long term experience
with perfluorobutane
in COMPASS RICH}

\author[label3,label2]{F.Bradamante}
\author[label3,label2]{A. Bressan}  
\author[label4,label2]{A. Cicuttin}
\author[label4,label2]{M. L. Crespo}
\author[label2]{C. Chatterjee}   
\author[label2]{P. Ciliberti}   
\author[label2]{S. Dalla Torre\corref{cor}}	
\ead{silvia.dallatorre@ts.infn.it}
\author[label4,label2]{W. Florian}
\author[label4,label3,label2]{L. Garc\'ia Ord\'o\~nez}
\author[label2]{M. Gregori}		
\author[label3,label2]{A. Kerbizi}
\author[label2]{S. Levorato}		
\author[label3,label2]{A. Martin}	
\author[label2]{G. Menon}	
\author[label4,label3,label2]{R. S. Molina }
\author[label3,label2]{A. Moretti}
\author[label2]{F. Tessarotto}	
\author[label4,label2]{Triloki}
\author[label4,label3,label2]{B. Valinoti}

\address[label2]{INFN Sezione di Trieste, Trieste, Italy}
\address[label3]{University of Trieste, Trieste, Italy}
\address[label4]{Abdus Salam ICTP, Trieste, Italy}
\cortext[cor]{Corresponding author}

\begin{abstract}
COMPASS RICH-1 has used  high-purity perfluorobutane as radiator gas since 2001. The operation and control of the radiator gas has evolved over years with continuous improvements.
We report on the experience gained in the 20 year-long operation of perfluorobutane as COMPASS RICH radiator.
\par
Very accurate values for the radiator gas refractive index are needed for high-performance particle identification. The procedure has evolved over years and the one presently in use, which provides refractive index estimate at the 1~ppm  level, is discussed. 
\par
Perfluorobutane procurement is becoming challenging, and the minimization of material waste is now a priority for the protection of the environment. Commercially available perfluorobutane needs dedicated filtering before usage and typical material losses in the filtering procedure were around 30\%. Recent efforts allowed us to reduce them to about 5\%.
A potential alternative to fluorocarbon radiators in gaseous RICHes is also presented.

\end{abstract}


\end{frontmatter}

\section{Introduction}
\label{sec:introduction}
RICHes with gaseous radiators still represent the only option for hadron Particle IDentification (PID) at high momenta. Fluorocarbon gasses are privileged because they offer high Cherenkov photon yield
and low chromaticity. These two characteristics 
are both relevant  to obtain fine resolution in the Cherenkov angle measurement and, therefore, in determining the momentum upper limit for effective PID.
\par 
We gained a 20 year-long experience using perfluorobutane (C$_4$F$_{10}$) in COMPASS RICH. We present items of general interest related to this long-term experience and future perspectives for gaseous RICHes.
\section{COMPASS RICH}
\label{compass-rich}
COMPASS RICH~\cite{compass-rich-silvia} is a wide acceptance Cherenkov imaging counter with gas
radiator, where single photon detection is performed by gaseous detectors equipped with CsI photocathodes detecting photons in the wavelength range 165-200~nm, and
multianode photomultiplier tubes, with sensitivity range in the visible domain extended till $\sim$200~nm. It accomplishes high-momentum hadron
identification in the COMPASS spectrometer at CERN SPS, which is dedicated to hadron physics studies. The counter, designed and built in the  last years of the twentieth century, has been in operation since 2001. It has  been upgraded several times over the 20 years of operation  to improve matching the demanding goals of the COMPASS physics programme and the continuously increasing trigger rate.
\par
The RICH has adopted the state-of-the-art technologies enriched with  original complements for radiator gas handling, the ultraviolet mirror wall, the multiwire proportional chambers equipped with CsI photocathodes, the multianode photomultiplier tubes coupled to telescopes of fused silica lenses  and the read-out electronics. It is the first RICH counter where, in a recent
upgrade, part of the original gaseous photon detectors have been replaced  with micropattern gaseous detectors designed for single photon detection, thanks to an original development.
\section{The radiator gas of COMPASS RICH-1 and the dedicated system}
\label{gas_system}
Perfluorobutane has been selected according to the general considerations illustrated in Sec.~\ref{sec:introduction}. Its
basic properties are provided in Table~\ref{tab:c4f10}. C$_4$F$_{10}$ has been used as radiator gas at CERN since the
days of the DELPHI experiment. Therefore, a wide experience was already
available at CERN and this has been largely beneficial for the use in the COMPASS
RICH.

\begin{table}
\caption{\emph{Perfluorobutane properties.
               }
               \label{tab:c4f10}
               }
\begin{center}
\begin{tabular}{|l|c|}
\hline  
	property	& value	\\
	\hline
	\hline
Molar mass 	& 238.028 g~$\times$~mol$^{-1}$\\
\hline
Density at &	11.21 kg/m$^3$ (gas)\\
101.3kPa boiling point & 1594 kg/m$^3$ (liquid)\\

\hline
Melting point &	-128~$^o$C\\
\hline
Boiling point 	& -1.7~$^o$C \\
\hline  
	\end{tabular}
	\end{center}
	\end{table}
\par
The radiator gas in the RICH vessel is controlled by the gas radiator system~\cite{gas_system}.
The system is based on a design already adopted to
 ensure perfluorobutane circulation in other RICH detectors~\cite{bosteels}, as those used in
HERA-B at HERA, DESY and in the space-born experiment CAPRICE. Its
 main tasks are providing, during detector operation, well controlled pressure
 conditions in the 80 m$^3$ RICH vessel; removing oxygen and water vapour
contaminants, in order to prevent building up impurities due to leaks; performing vessel filling with perfluorobutane before a data taking period and
radiator gas recovery at the end of the period. The recovery operation ensures C$_4$F$_{10}$ storage in a dedicated tank for future usage. For these purposes,
the system main components are two oil-free compressors, working in parallel, which continuously extracts gas from the vessel at constant rate in order to ensure the gas circulation, a pressure sensor installed on top of the radiator vessel for continuous monitoring of the internal relative pressure, a pneumatic valve on the input line, which controls the input gas flow to preserve the relative pressure inside the vessel. In fact, the pressure regulation
 is obtained by varying the gas flow at the input to the vessel. Oxygen and water
 vapour traces are removed by filtering cartridges with 5A molecular sieves
and Cu-catalyst, which are permanently in series in the circulation system.
 The vessel is flushed with nitrogen during the shutdown periods. N$_2$ and
 C$_4$F$_{10}$ separation during filling and recovery is based on the different boiling
 points of the two gases: distillation is performed at -35$^o$C. The pressure in
 the separator section is kept at 400-500~kPa: nitrogen vented out thus contains residual C$_4$F$_{10}$ at the 4-5~\% level. The control of the whole system is  performed via a Programmable Logic Control (PLC).  The gas system is complemented by monitoring instrumentation, including commercial instrumentation as a hygrometer, an oxygenmeter and a binary gas analyser and custom devices: a sonar system to determine the gas
composition by measuring the speed of the sound in the gas~\cite{sonar} and a setup
 for gas transparency measurements, shortly described here. A monochromatic light beam is obtained coupling a deuterium lamp to a monochromator
and detected by two PhotoMultiplier Tubes (PMT) with wavelength shifters.
The monochromatic light beam enters a measuring chamber filled with the
 radiator gas, where the light beam is split with a beam splitter between the
light detected by the reference PMT and the light detected by the measuring PMT, both read-out via picoammeters. The system can measure in the
wavelength range 154-220~nm and the transmission through a path of 187~cm
is measured, this being the difference in length between the paths of the light
reaching the two PMTs. The estimated wavelength spread is 0.5~nm and the estimated overall absolute calibration error is $<$0.25~nm. The system is calibrated at each measurement collecting reference spectra with nitrogen. The
error associated to the transmission measurements, estimated from repeated measurements, is at the 1\% level.
\par
\begin{figure}
\begin{center}
\includegraphics[width=0.35
\textwidth]{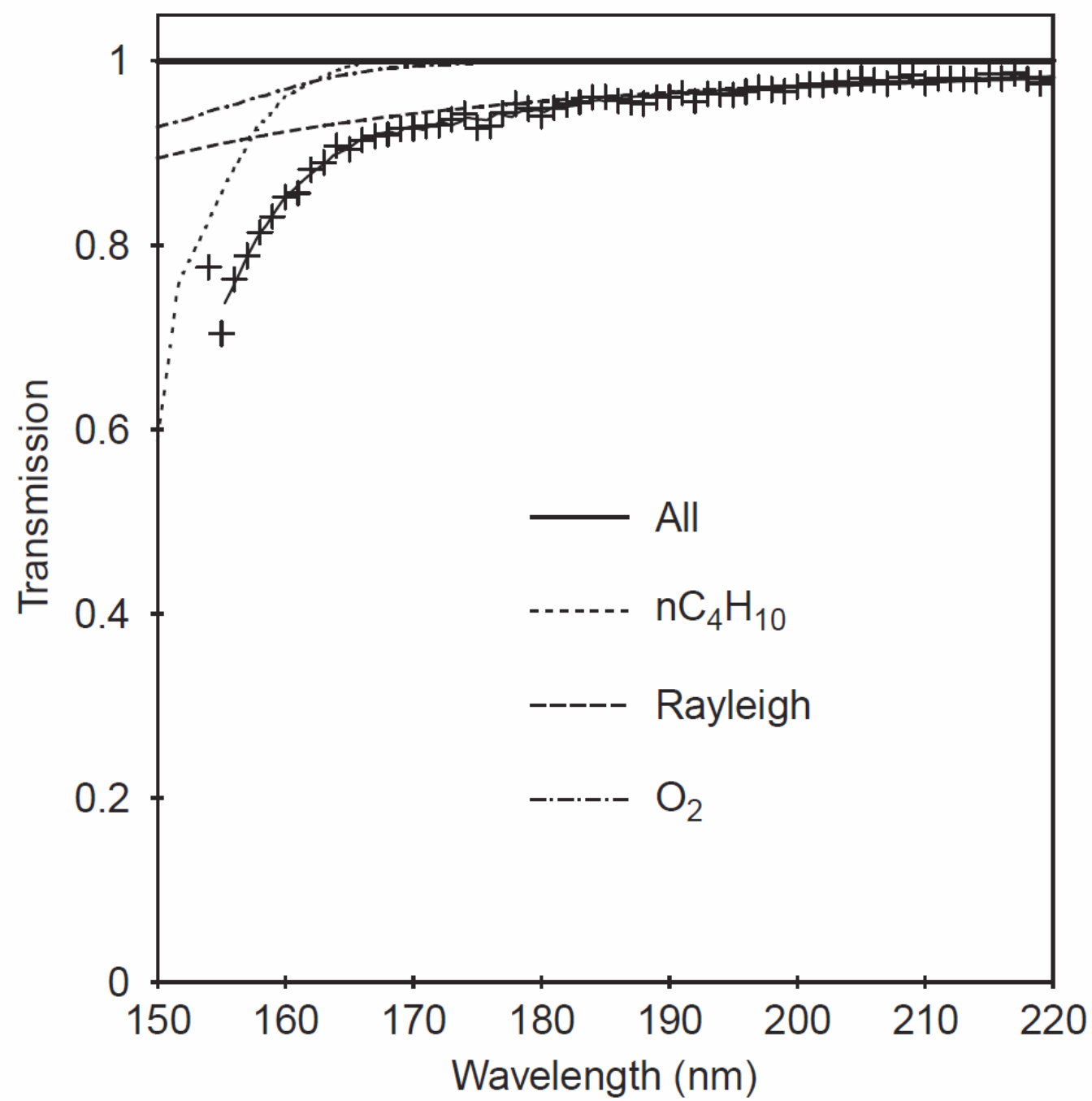}
\caption{ Typical UV light transmission through 1.87 m of 
radiator gas, as measured on-line during the data taking; 
the fitted curve allows to disentangle the main contributions
to UV light absorption, namely: Rayleigh scattering, oxygen,
water vapour, hydrocarbons and alkene traces; the various
relevant contributions are also plotted: they give indications
on the amount of residual contaminants; in particular water
vapour traces below 1 ppm and oxygen traces below 3 ppm are
routinely obtained.
\label{fig:gas-transparency} 
}
\end{center}
\end{figure}

The most relevant parameters describing the overall performance of the
 radiator gas system are a typical gas flow rate in the vessel of 3-5 m$^3$/h, a
typical oxygen and water vapour contamination at a few ppm level, residual
nitrogen dissolved in the perfluorobutane atmosphere in the vessel in the range
3-4\%, relative pressure stable over months within 100 Pa and a typical gas
transparency as shown in Fig.~\ref{fig:gas-transparency}.
The relatively slow gas flow rate and the large vessel volume can result
in the formation of a temperature field in the radiator volume. This effect
is prevented thanks to the fast gas circulation, a parallel circulation system
by a turbo-pump that provides gas flow at a rate of
20~m$^3$/h. It also contributes to reduce the overall temperature evolution of
the radiator gas thanks to the use of heat exchangers.
The radiator gas system is stable, robust and easy to operate. It exhibits  occasional instabilities, typically once or twice per year of operation, in case
of violent thunderstorms, when atmospheric pressure increases up to 3000 Pa
in less than 30 minutes have been observed. The circulation system is not
able to provide in real time the required gas amount for compensation and an error condition is detected, causing a system halt. Recovery is by manual
intervention. Regular operation is always restored in less than half an hour.

\section{Perfluorobutane provider, purity and transparency in the ultraviolet domain}
\label{sec:transparency}
Perfluorobutane is intrinsically transparent in the UV region. Nevertheless, the material from commercial providers is not 100\% pure and
pollutants can severely affect its transparency, in particular in the UV domain. Procedures to remove the non-UV transparent contaminants have been
developed~\cite{gas_cleaning}. The success of these approaches depends on the nature of
the impurities. Among non-UV transparent pollutants, oxygen, water vapour
and alkenes are easier to remove, while aromatics can be removed only by
lengthy procedures resulting in important material losses.
Historically, the perfluorobutane used in  the CERN experiments is 3M\footnote{Minnesota Mining and Manufacturing Company, 3M Center, St. Paul, MN, USA} performance fluid PF-5040. The specifications guarantee 99\% C$_4$F$_{10}$ in mole. The majority  of
the 3M batches, but not all of them,  could be made UV transparent. 3M is no longer producing the performance fluid PF-5040. For the COMPASS 2021 run, a new
provider has been identified:  Perfluorobutane Technical-grade by
F2 Chemicals\footnote{F2 Chemicals LTD, Lea Lane, Lea Town, Preston, Lancashire, PR4 ORZ, UK} with 98\% C$_4$F$_{10}$ in volume according to specifications.
Samples by F2 Chemicals have been made fully transparent in the UV domain of interest
with the procedures already used for the 3M material.
\par
At COMPASS, a dedicated cleaning gas system, in operation since 2002,
removes the pollutants of the raw material before its use as radiator gas.
Perfluorobutane is circulated in closed loop in liquid phase through activated
carbon and 13X molecular sieve, later replaced by 3A molecular sieve. Oxygen is removed in a cool section (T:~-60$^o$C) where C$_4$F$_{10}$ condensates and the
liquid drops return to the bottle, while the gaseous component is vented out.
The typical material loss is between 20\% and 50\% for the different batches, related to the different level of contaminants.
\par
The gas cleaning system has been modified in 2021 with the goal of reducing the material loss in the cleaning procedure.
For this purpose, two main modifications have been introduced. A section with pre-filtering in gas phase has been added with cartridges of 3A molecular sieve, activated carbon filter and Cu-catalyst.
They are quickly saturated when filtering the raw material trapping it in gas phase. Thanks to the gas phase, the cartridge regeneration process causes the loss of a limited amount of material. Moreover, the use of Cu-catalyst to remove oxygen allows to skip the low-efficient oxygen separation by distillation.
The improvement is remarkable resulting in a material loss of 5\% only.
\section{The radiator refractive index}
\label{sec:refractive_index}
\begin{figure}
\begin{center}
\includegraphics[width=0.40\textwidth]{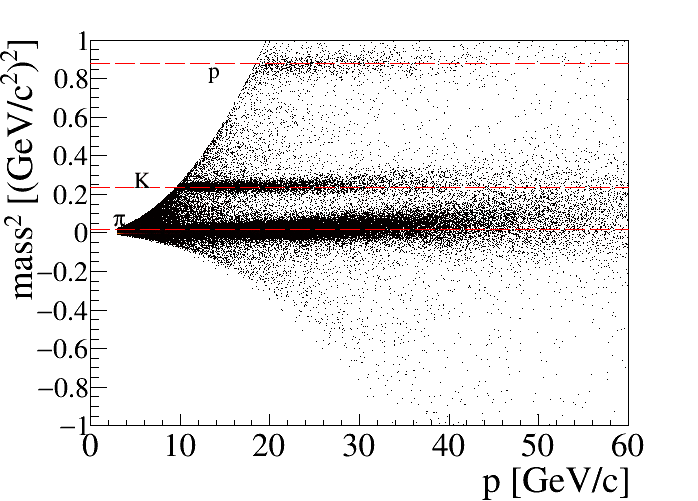}
\end{center}
\caption{Squared mass from equation~\ref{eq:m2} versus particle momentum.
\label{fig:m2} 
}
\end{figure}

\begin{figure}
\begin{center}
\includegraphics[width=0.40\textwidth]{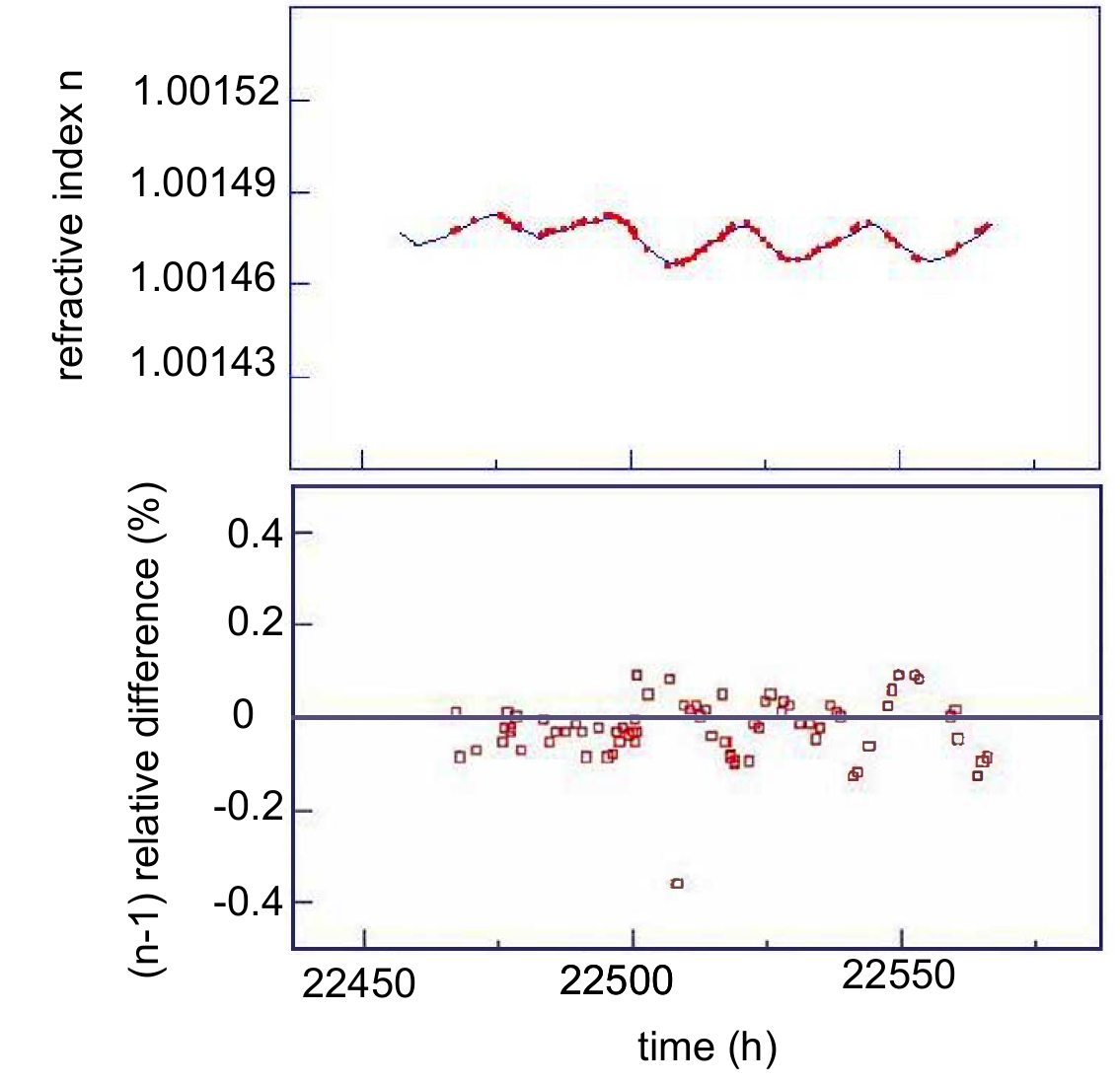}
\end{center}
\caption{Refractive index for the gaseous photon detectors. 
Top: The measured refractive index n versus time (open squares) and the 
refractive index calculated from the temperature and pressure evolution (continuous line). Bottom: the relative difference of (n-1) from the measurements and from the calculation. The time scale refer to the COMPASS experiment clock including all the data taking periods collected over years.  
\label{fig:n1bis} 
}
\end{figure}

A key ingredient for PID is the precise knowledge 
of the effective refractive index n, that can only be extracted from the data
collected with the RICH. In fact, it depends on many parameters,
mainly on the purity of the gas, the pressure and the 
temperature. Thus its value is time dependent. 
For each photon associated to a particle of measured momentum, n
is computed from the Cherenkov equation, assuming
the $\pi$ mass. The distribution of such values exhibits a Gaussian-like peak
over a low-level background. The mean value of the Gaussian best fit of the peak is assumed as the
index value. 
Even if $\pi$ are the most abundant hadrons in the final state, other particle species are present and they can distort the peak. Therefore, only high momentum particles (40-100 GeV/c) are selected, for which the Cherenkov angle in perfluorobutane is equal to the maximum within the measuring error. 
The correct determination of the refractive index can be verified from the data themselves using the equation: 
\begin{equation}
\label{eq:m2}
m^2=p^2\frac{2(n-1)-\theta}{1-(n-1)}
\end{equation}
where m is the particle mass, p its momentum and $\theta$ the measured Cherenkov angle.
m$^2$ calculated according to equation~\ref{eq:m2} is not dependent on p when n-value is correctly determined (Fig.~\ref{fig:m2}).
The refractive index must be known 
with high accuracy: a $\Delta$n of 10$^{-5}$ with respect to a reference value of 
n~=~1.00150, namely a variation of the order of 1\% in (n-1),  corresponds to a  variation of the Cherenkov angle of the order of 0.2~mrad.
Temperature and pressure oscillations causing (n-1) variations at 
the 2\% level or more are regularly observed, even induced by the day and night 
excursion of the environmental parameters. Sizable changes of the refractive index are observed every hour. They require a 
continuous adjustment of the n-value for data processing.
A double data reconstruction, once to extract n  and a second one to process 
the data with the correct n-value cannot be afforded.
In order to follow the refractive index evolution versus time extracting 
it from the data, a sample of data collected every hour should be analyzed.
This corresponds  to the need  of analyzing twice approximately
10\% of the collected data. The load is heavy, taking into account 
that COMPASS typically collects  a few  PB of data per day.
Therefore, n is measured 
from a limited number of data samples 
and the evolution of its value is calculated from the measured temperature and pressure parameters.
This approach provides (n-1)-values accurate at the per mil level
(Fig.~\ref{fig:n1bis}).
\section{Future perspectives for the radiator gasses for RICH counters}
Any potential alternative to the use of fluorocarbons as gas radiators in Cherenkov imaging application should match the following requirements: similar refractive index, in the high-value range for gasses, in order to guarantee a good rate of produced Cherenkov photons, and limited chromaticity
to preserve the high resolution, being the chromatic dispersion the irreducible component of the Cherenkov angle dispersion. 
An option is offered by pressurized noble gasses, which have limited chromaticity. Helium and neon would require severe high pressure to obtain refractive indexes in the fluorocarbon range. Argon and xenon can mimic fluorocarbon refractive indexes at moderate pressure (Table~\ref{tab:refrective-indexes}). 
The chromatic dispersion varies with the quantum efficiency spectrum of the photosensors. The chromatic dispersion of three fluorocarbons, of Ar and of Xe assuming three photosensors with different spectral response are provided in Table~\ref{tab:chromaticity}.  
Ar chromatic dispersion is similar to that of fluorocarbons, in particular to C$_4$F$_{10}$ one both in the visible and UV domain. Xe chromatic dispersion is acceptable in the visible and near UV domain, offering the advantage that lower gas pressure is required to mimic the fluorocarbons.
\par
The above analysis indicates that pressurized noble gasses are adequate radiators for gaseous RICHes. The gas choice has to be optimized according to the selected photosensors and their quantum efficiency spectrum. This option is under consideration for the forward RICH of the ePIC experiment at the US Electron-Ion Collider (EIC).
\begin{table}
\caption{\emph{Ar and Xe absolute pressure to be applied to mimic the refractive indexes of three fluorocarbons typically used in gaseous RICHes. Fluorocarbon are assumed at atmospheric pressure.}
               \label{tab:refrective-indexes}
               }
\begin{center}
\begin{tabular}{|l|c|c|}
\hline  
	Fluorocarbon	& Ar, Pressure (bar)	& Xe, Pressure (bar)\\
	\hline
	\hline
CF$_4$ 	& 1.7 & \\
\hline
C$_2$F$_6$ 	& 2.9 & 1.2\\
\hline
C$_4$F$_{10}$ & 4.6	& 1.9 \\
\hline
	\end{tabular}
	\end{center}
	\end{table}
\begin{table}
\caption{\emph{Chromatic dispersion of three fluorocarbons, of Ar and of Xe assuming the following photosensors: MAPMT type R7600-03-M16 by Hamamatsu, with UV extended glass window,; SiPM-14520 by Hamamatsu; SiPM-13615 by Hamamatsu. }
               \label{tab:chromaticity}
               }
\begin{center}
\begin{tabular}{|l|c|c|c|}
\hline
    photosensor	&  	MAPMT 	& SiPM-14520 	& SiPM-13615\\
    \hline
   Wavelength	&   & &\\
   range (nm)	&  200-700 & 270-900 & 320-900\\
   \hline
	\hline 
	$\sigma_{\theta}$/$\theta$ (CF$_4$) & 2.3 & 1.2 & 0.8 \\
	$\sigma_{\theta}$/$\theta$ (C$_2$F$_6$) & 2.5 & 1.3 & 0.9 \\
	$\sigma_{\theta}$/$\theta$ (C$_4$F$_{10}$) & 3.3 & 1.7 & 1.1 \\
	$\sigma_{\theta}$/$\theta$ (Ar) & 3.3 & 1.7 & 1.5 \\
	$\sigma_{\theta}$/$\theta$ (Xe) & 7.9 & 3.2 & 2.3 \\
\hline
	\end{tabular}
	\end{center}
	\end{table}
\section{Conclusions}
\label{conclusions}
We have presented key elements of our long-term experience with perfluorobutane used as gas radiator in COMPASS RICH.
\par
The radiator gas system has been described as well as its performance over the more than 20-year usage at COMPASS. The system is reliable
showing some criticality in case of occasional fast variation of the atmospheric pressure, as in case of thunderstorms. The key aspects of purity and transparency has been discussed together with procedures that allow to remove the UV opaque pollutants. The material loss due to the gas filtering has been reduced thanks to the experience gained over time and, recently, a rate of 5\% losses has been obtained.
\par
The determination and monitoring of the gas refractive index has been presented, together with a cross-check of its correct value based on basic physical quantities.
\par
Fluorocarbons are characterized by huge-values of their Global Warming Power (GWP). This fact poses ethical and procurement issues for their use in future devices. The usage of pressurized noble gasses as alternatives to fluorocarbons has been discussed with promising perspectives.


\section*{Acknowledgments}
One of the authors (C.~C.) is supported by  the European Union’s Horizon 2020 Research and Innovation programme under Grant Agreement AIDAinnova - No 101004761.


\end{document}